\documentclass[twocolumn,showpacs,preprintnumbers,amsmath,amssymb,nofootinbib]{revtex4-1}
\usepackage{graphicx}
\usepackage{dcolumn}
\usepackage{bm}
\usepackage{color}

\begin{document}

\title{Hybrid Stars in an SU(3) Parity Doublet Model}

\author{V.~Dexheimer}
 \email{vdexheim@kent.edu}
\affiliation{Department of Physics, Kent State University, Kent OH, USA}

\author{J.~Steinheimer}
 \email{JSFroschauer@lbl.gov}
\affiliation{Lawrence Berkeley National Laboratory, 1 Cyclotron Road, Berkeley, CA 94720, USA}
\affiliation{FIAS, Johann Wolfgang Goethe University, Frankfurt am Main, Germany}

\author{R. Negreiros}
 \email{negreiros@fias.uni-frankfurt.de}
\affiliation{Physics Department, Universidade Federal Fluminense,
Niter\'oi, Brazil}
\affiliation{FIAS, Johann Wolfgang Goethe University,
Frankfurt am Main, Germany}

\author{S. Schramm}
\email{schramm@fias.uni-frankfurt.de}
\affiliation{FIAS, Johann Wolfgang Goethe University,
Frankfurt am Main, Germany}

\date{\today}
\begin{abstract}
We apply an extended version of the SU(3) parity model, containing quark degrees
of freedom, to study neutron stars. The model successfully reproduces the main
thermodynamic features of QCD which allows us to describe the composition of
dense matter. Chiral symmetry restoration is realized inside the
star and the chiral partners of the baryons appear, their masses becoming
degenerate. Furthermore, quark degrees of freedom appear in a transition to a
deconfined state. Performing an investigation of the macroscopic properties of
neutron stars, we show that observational constraints, like mass and thermal
evolution, are satisfied and new predictions can be made.
\end{abstract}

\maketitle
\section{Introduction}
The study of strong interaction physics under extreme conditions is a central
topic of nuclear physics with a large number of experimental and theoretical programs
focusing on the area. These conditions comprise large
temperatures, densities, as well as extreme values of nuclear isospin. In
ultra-relativistic heavy-ion collisions a very hot fireball is created in the
collision zone and, at high temperatures, hadronic matter is assumed to melt into its
constituents, quarks and
gluons. The net baryon density in such reactions is determined by the beam energy. At
the planned energies in the upcoming
FAIR (Facility for Antiproton and Ion Research) at GSI, a hot and relatively dense system will be produced. A
central point of these investigations is the understanding of the quark-hadron
phase transition. However, while relatively high densities and finite temperatures might
be reached in laboratory experiments, the study of neutron stars is essential if
one wants to probe the low temperature and high density region of the phase
diagram of strongly interacting matter.

Observations of neutron star masses (and possibly radii) provide the standard way of constraining the inner composition of these objects. More precisely, recent observations have set new high mass constraints for neutron stars
(PSR J1614-2230 \cite{Demorest:2010bx} being the most important), and equations of state aimed to describe compact stars are thus expected to provide objects with high masses 
\cite{arXiv:1006.5660,arXiv:1107.2497,arXiv:1112.0234,Djapo:2008au,Whittenburyetal.(2012)}. In the
case of hybrid stars (neutron stars with a quark core in its center), a quark
phase based on a simple non-interacting quark model like the MIT bag model
tends to reduce the maximum mass significantly (see the discussion in
\cite{arXiv:1011.2233}). A quark phase that includes strong repulsive
interactions, however, may have an equation of state quite similar to a
purely hadronic one. This prevents the softening of the matter and the drop
in maximum mass \cite{nucl-th/0411016,arXiv:1102.2869,arXiv:1108.0559}.
It has been found, however, that models with a strong
repulsive quark-quark
interaction make the description of lattice results at $\mu_B=0$ nearly impossible \cite{Kunihiro:1991qu,Ferroni:2010xf,Steinheimer:2010sp}, since
including a repulsive
mean field interaction strongly decreases the quark number susceptibility.
Therefore, most current successful hybrid star models would fail in this regard, if they were to be applied to the high-temperature, low-density regime.

At high temperatures, QCD exhibits a crossover to a deconfined phase and
the quarks and gluons become the dominant degrees of freedom \cite{Aoki:2006we}. The temperature at which
this transition takes place is
estimated to be  $T_{dec}\approx 150 - 160$ MeV
\cite{Borsanyi:2010cj,Bazavov:2010sb}. A phase transition is also expected to
take place at high densities, where baryons start to overlap, and quarks and gluons
become the effective degrees of freedom. This indicates that at some point a
hadronic model will not be able to appropriately describe the matter present
inside a neutron star, and a deconfinement mechanism needs to be introduced.

In the following, we will discuss a theoretical approach that is able to describe
the conditions found in compact stars
as well as those created in heavy-ion collisions. The aim of this approach is to
find a unified description for the thermodynamic properties of QCD which can be
applied to compact stars and heavy ion collisions at different beam
energies while being in accordance with lattice QCD results at vanishing
baryon density. 
An approach of this type is essential if one wants to investigate a phase diagram that has a region of a cross-over
transition to quark matter (as it is clearly established by lattice QCD calculations) and a first-order transition at high densities and low temperatures.
This is not possible by combining separate models for the hadronic and quark phase. 
Such a first-order transition at low temperature has also been observed in recent lattice monte-carlo calculations of an effective QCD Lagrangian \cite{philipsen}.
In addition, a model that can cover the physics at high temperature as well as at high density can serve as an important tool for the increasingly important studies of black hole formation and black-hole neutron-star mergers, where temperatures up to 90 MeV and high densities might be reached \cite{arXiv:1102.3753,shen}.

In a previous paper, an extended quark $\rm SU(3)_f$ parity doublet
model was introduced for this purpouse \cite{Steinheimer:2011ea}. 
In the parity model \cite{Detar:1988kn,Hatsuda:1988mv}, the mass splitting
between the nucleons and the respective chiral partners is generated by the
spontaneous breaking of chiral symmetry and their coupling to the corresponding
order parameter, the scalar (sigma) field. The same applies for
the baryon octet in the $\rm SU(3)_f$ case \cite{Nemoto:1998um,Jido:1999hd}.
When chiral symmetry is restored and the sigma field vanishes, the chiral partner masses become degenerate. A good
description of nuclear saturation properties, as well as neutron star
observables can easily be achieved within this formalism
\cite{Dexheimer:2007tn,Dexheimer:2008cv}. In Ref. \cite{Sasaki:2010bp} a phase
diagram for chiral symmetry restoration was calculated using an $\rm SU(2)_f$
version of the parity model. 

In addition, the parity model has been shown to describe the lattice results at
$\mu_B = 0$ and can be used in dynamical models for
relativistic heavy ion collisions \cite{Steinheimer:2009nn,Santini:2011zw,Petersen:2011sb}. We
will apply this model to the study of neutron stars. A complete phase diagram
for iso-spin symmetric matter, using the extended version of the $ \rm SU(3)_f$
parity model, which contains quark degrees of freedom, has been calculated in
\cite{Steinheimer:2011ea}. As expected, at low temperatures the nuclear matter
liquid-gas phase transition and the chiral symmetry
restoration are of first order. Within this approach the deconfinement phase transition, on the other hand, is a
crossover. Furthermore, chiral symmetry restoration and deconfinement do not
coincide, and at intermediate densities the matter is chirally symmetric but
still confined. Whether such a chirally symmetric hadronic phase can be the
$N_c=3$ equivalent of the $N_c=\infty$ quarkyonic phase \cite{McLerran:2007qj}
is still subject to debate
\cite{Lottini:2011zp,Bonanno:2011yr,Giacosa:2011uk}. 
 
In this paper we investigate the properties of electrically-neutral chemically-equilibrated matter in the framework of the model described above. We will
investigate the influence of the chiral partners, hyperons and quark matter on
the macroscopic properties of a neutron star, such as its gravitational mass,
radius, and thermal evolution.

\section{The SU(3) Parity Doublet Model}

Considering that the baryons are grouped in doublets ($B_+$ and $B_-$), in which
they belong to the same multiplet, the components of the fields $\psi_1$ and
$\psi_2$ transform under $L$ and $R$ rotations like
\begin{eqnarray}
&\psi'_{1R}  =  R \psi_{1R}, &~ ~ ~ ~ ~ \psi'_{1L} = L \psi_{1L}~, \ \nonumber
\\ 
&\psi'_{2R} = L \psi_{2R}, &~ ~ ~ ~ ~ \psi'_{2L} = R \psi_{2L}~.
\end{eqnarray}
This formalism allows the presence of a bare mass term in the Lagrangian
density that does not break chiral symmetry
\begin{eqnarray}
&m_{0}( \bar{\psi}_2 \gamma_{5} \psi_1 - \bar{\psi}_1 \gamma_{5} \psi_2 ) = 
\nonumber \\
&m_0 (\bar{\psi}_{2L} \psi_{1R} - \bar{\psi}_{2R} \psi_{1L} - \bar{\psi}_{1L}
\psi_{2R} + \bar{\psi}_{1R} \psi_{2L})~,
\end{eqnarray}
where $m_0$ represents a mass parameter. Since the term proportional to $m_0$ mixes
the upper and lower components of the parity doublets, one diagonalizes the
matrix by introducing new fields $B$ with a diagonal mass matrix. Keeping
only the diagonal meson contributions, the scalar and vector condensates in the
mean field approximation, the resulting Lagrangian reads
\begin{eqnarray}
{\cal L_B} &=& \sum_i (\bar{B_i} i {\partial\!\!\!/} B_i)
+ \sum_i  \left(\bar{B_i} m^*_i B_i \right) \nonumber \\ &+&
\sum_i  \left(\bar{B_i} \gamma_\mu (g_{\omega i} \omega^\mu +
g_{\rho i} \rho^\mu + g_{\phi i} \phi^\mu) B_i \right)~,
\label{lagrangian2}
\end{eqnarray}
where the coupling constants for the baryons with the mesons $\omega$, $\rho$
and the strange meson $\phi$ are $g_{B\omega}$, $g_{B\rho},$ and $g_{B\phi}$, respectively.
We do not
include a vector-meson self-interaction to avoid an extra softening of the equation of state, which in turn would yield
neutron stars with very low maximum mass
\cite{Zschiesche:2006zj,Dexheimer:2008cv}. This happens because as the coupling
constant increases, the respective vector-isoscalar field decreases (in order
to reproduce saturation properties), thus making the repulsive part of the
strong force less pronounced.

The expression for the effective masses of the baryons for
isospin symmetric matter reads
\begin{eqnarray}
m^*_i &=& \sqrt{ \left[ (g^{(1)}_{\sigma i} \sigma + g^{(1)}_{\zeta i}  \zeta
)^2 + (m_0+n_s m_s)^2 \right]}\nonumber \\
&\pm& g^{(2)}_{\sigma i} \sigma \pm g^{(2)}_{\zeta i} \zeta~. \label{mef}
\end{eqnarray}
In the above equation, the
term that does not correspond to the bare mass is generated by the scalar mesons
$\sigma$, the strange $\zeta$, and by the SU(3) breaking mass term with $m_s = 150$ MeV.
The last term is responsible for the generation of an explicit mass corresponding to
the strangeness $n_s$ of the baryon. 
The values of the various coupling constants of the  baryons with the meson fields are given in Ref. \cite{Steinheimer:2011ea}. 
The couplings $g^{(2)}$ are chosen in order
to reproduce the splitting of the masses of the parity partners in vacuum,
further assuming, for simplicity, an equal splitting of the masses of the
various baryonic doublets ($g_{\sigma i}^{(2)}=-0.850$, $g_{\zeta i}^{(2)}=0$). The coupling constants $g_i^{(1)}$ are chosen in order to reproduce vacuum masses of
baryons. The bare mass parameter is set to $m_0=810$ MeV. Such a high value is
necessary in order to reproduce reasonably massive stars while keeping the
compressibility of saturated matter at reasonable values
\cite{Dexheimer:2007tn,Dexheimer:2008cv}. The value for the strange quark mass
$m_s$ is fixed to $150$ MeV.

The nucleonic vector interactions are tuned to reproduce reasonable values for the nuclear
ground state properties (binding energy per baryon of $\sim-16$ MeV and baryonic
density of $\rho_0 = 0.15\ \rm{fm}^{3}$ at saturation) while the hyperonic vector interactions are
tuned to generate reasonable optical potentials for the hyperons in ground state
nuclear matter ($U_\Lambda \sim -28$ MeV, $U_\Sigma \sim  -14.5$ MeV and $U_\Xi \sim  -18$ MeV).
The equations of motion are obtained by minimizing the grand canonical
potential (a function of baryonic chemical potential and temperature), and are
solved self-consistently in mean field approximation. For non-zero temperature
calculations, a hadronic heat bath is included.

We choose the vacuum mass of the chiral parter for the nucleons to be $1535$ MeV, the lighter resonance with spin $1/2$ and negative parity. Lower values for this mass ($\sim 1200$ MeV for instance) would lead to chiral restoration taking place at lower densities \cite{Zschiesche:2006zj}, and to less massive stars \cite{Dexheimer:2007tn}. In addition, using a formula for the width as a function of the bare mass parameter $m_0$ \cite{Detar:1988kn}, one finds that the width of the less massive resonance would be too small to have escaped experimental detection \cite{Dexheimer:2008cv}. Other candidates like the $1650$ MeV also have been suggested \cite{Gallas:2009qp}. For the hyperons, we simply chose the splitting between the particles and the respective parity partners to have the same value as for the nucleons, thus keeping the number of parameters to a minimum. This assumption agrees quite well with some parity partner candidates, like the $\Lambda(1670)$ and the $\Sigma(1750)$. In the case of the $\Xi$ the data is unclear. 

\section{Inclusion of Quarks}
Following \cite{Steinheimer:2011ea}, the effective masses of the quarks are
generated by the scalar mesons except for a small explicit mass term ($\delta
m_q=5$ MeV and $\delta m_s=150$ MeV for the strange quark) and $m_{0_q}$
\begin{eqnarray}
&m_{q}^*=g_{q\sigma}\sigma+\delta m_q + m_{0q}~,&\nonumber\\
&m_{s}^*=g_{s\zeta}\zeta+\delta m_s + m_{0q}~,&
\end{eqnarray}
with $g_{q\sigma}=g_{s\zeta}= 4.0$. As was the case for baryons,
we have also introduced a mass parameter $m_{0q}= 200$ MeV for the quarks. 

We use the Polyakov loop $\Phi$ as the order parameter for deconfinement. The
field $\Phi$ is defined via $\Phi=\frac13$Tr$[\exp{(i\int d\tau A_4)}]$, where
$A_4=iA_0$ is the temporal component of the SU(3) gauge
field \cite{Fukushima:2006uv,Allton:2002zi,Dumitru:2005ng}. The coupling of the
quarks to the Polyakov loop is introduced in the thermal energy of the quarks
(see \cite{Steinheimer:2011ea} for more details). All thermodynamical quantities
are derived from the grand canonical potential that includes the effective
potential $U(\Phi,\Phi^*,T)$, which controls the dynamics of the Polyakov-loop.
In our approach we adopt the ansatz proposed in
\cite{Ratti:2005jh,Fukushima:2003fw}
\begin{eqnarray}
U&=&-\frac12
a(T)\Phi\Phi^*\nonumber\\&+&b(T)ln[1-6\Phi\Phi^*+4(\Phi^3\Phi^{*3}
)-3(\Phi\Phi^*)^2]~,
\end{eqnarray}
with $a(T)=a_0 T^4+a_1 T_0 T^3+a_2 T_0^2 T^2$ and $b(T)=b_3 T_0^3 T$. The
parameters are fixed, as in Ref. \cite{Ratti:2005jh}, by demanding a first order
phase transition in the pure gauge sector at $T_0=270$ MeV, and that the
Stefan-Boltzmann limit of a gas of glouns is reached for $T \rightarrow \infty$.

\begin{figure}[t]
 \centering
\includegraphics[width=0.4\textwidth]{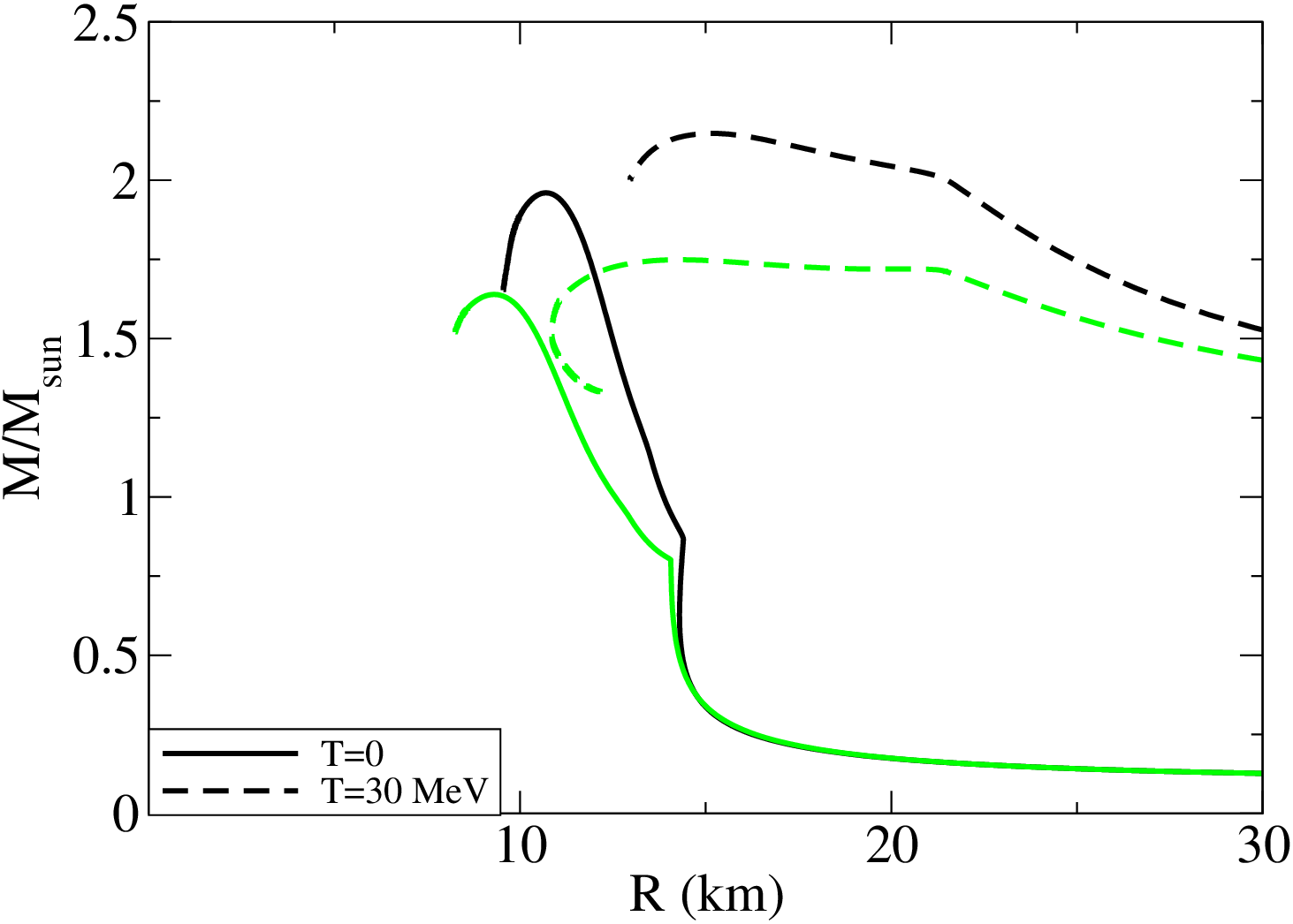}
 \caption{(Color online) Mass-radius diagram for star families with different
temperatures shown for Models A and B (black and green lines, respectively). For the
cold case we have included a cold crust from \cite{Baym:1971pw} and for the warm case we
have included a crust with $s/\rho_B=4$ in accordance with
\cite{Lattimer:1991nc}.}
 \label{mass}
\end{figure}

To remove the hadronic contributions from the equation of state at high
temperatures and densities, we introduce excluded volume effects. The inclusion
of finite-volume effects in thermodynamic models for hadronic matter was
proposed in
\cite{Hagedorn:1980kb,Baacke:1976jv,Gorenstein:1981fa,Hagedorn:1982qh,
Rischke:1991ke,Cleymans:1992jz,Kapusta:1982qd,Bugaev:2000wz,Bugaev:2008zz,
Satarov:2009zx,Hempel:2011kh,Lattimer:1991nc, Shen:1998gq}. In recent publications \cite{Steinheimer:2010ib,Steinheimer:2011ea} we adopted
this ansatz to successfully describe a smooth transition from a hadronic to a
quark dominated system. The modified chemical potential $\widetilde{\mu}_i$,
which is connected to the real chemical potential $\mu_i$ of the $i$-th particle
species, is obtained by the following relation
\begin{equation}
\widetilde{\mu}_i=\mu_i-v_{i} \ P~,
\end{equation}
where $v_{i}=1\ \rm{fm}^{3}$ is the excluded volume of a particle of species $i$
(zero for quarks), and $P$ is the sum over all partial pressures. To be
thermodynamically consistent, all densities (energy density $\widetilde{e_i}$,
baryon density $\widetilde{\rho_i}$ and entropy density $\widetilde{s_i}$) were multiplied by a volume correction factor $f$, which is the ratio of the
total volume $V$ and the reduced volume $V'$ not being occupied. As a
consequence, the chemical potentials of the hadrons are decreased by the quarks,
but not vice-versa. In other words, as the quarks start appearing, they
effectively suppress the hadrons by changing their chemical potential, while the
quarks are only affected through the volume correction factor $f$.

A shortcoming of the parametrization above is that the compressibility of bulk
matter at saturation density is too high ($525$ MeV) and the equations of state is a bit too stiff a low densities when compared with heavy ion collision data from Ref. \cite{Danielewicz:2002pu}. For this reason, in
addition to the parametrization discussed above (Model A hereafter), we use a
second parametrization (Model B) that leads to more realistic values for the
compressibility ($\kappa \approx 330$ MeV). The values of the modified
parameters in Model B are $v_i=0.5\ \rm{fm}^{3}$, $g_{N\omega}=5.02$,
$g_{\sigma}^{(1)}=-7.79$ and $g_{\sigma}^{(2)}=-0.76$ (see Table I of Ref. \cite{Steinheimer:2011ea} for the entire parameter set). The symmetry energy obtained at saturation is $32.6$ and $32.4$ MeV and its slope $L$ is $99$ and $92$ MeV, respectively for Model A and B. Both of these are in reasonable agreement with symmetry energy constraints from Ref. \cite{Lattimer:2012xj}.

It is important to notice that, although $v_i$ is directed related to the position of the phase transitions that occur when new particles appear in the system, it cannot be varied freely. This parameter is chosen mainly to reproduce lattice QCD results at $\mu_B=0$ for Model A. For Model B, one has more freedom for the choice of $v_i$, but the phase transitions reproduced do not change qualitatively with this variation. This happens because in this type of unified equation of state, the order in which the particles appear depends strongly on charge neutrality. For example, negatively charged particles, like the down quark, will always be favored with respect to positive ones.

\section{Results}

\begin{figure}[t]
 \centering
\includegraphics[width=0.4\textwidth,clip,trim=8 0 0 0 ]{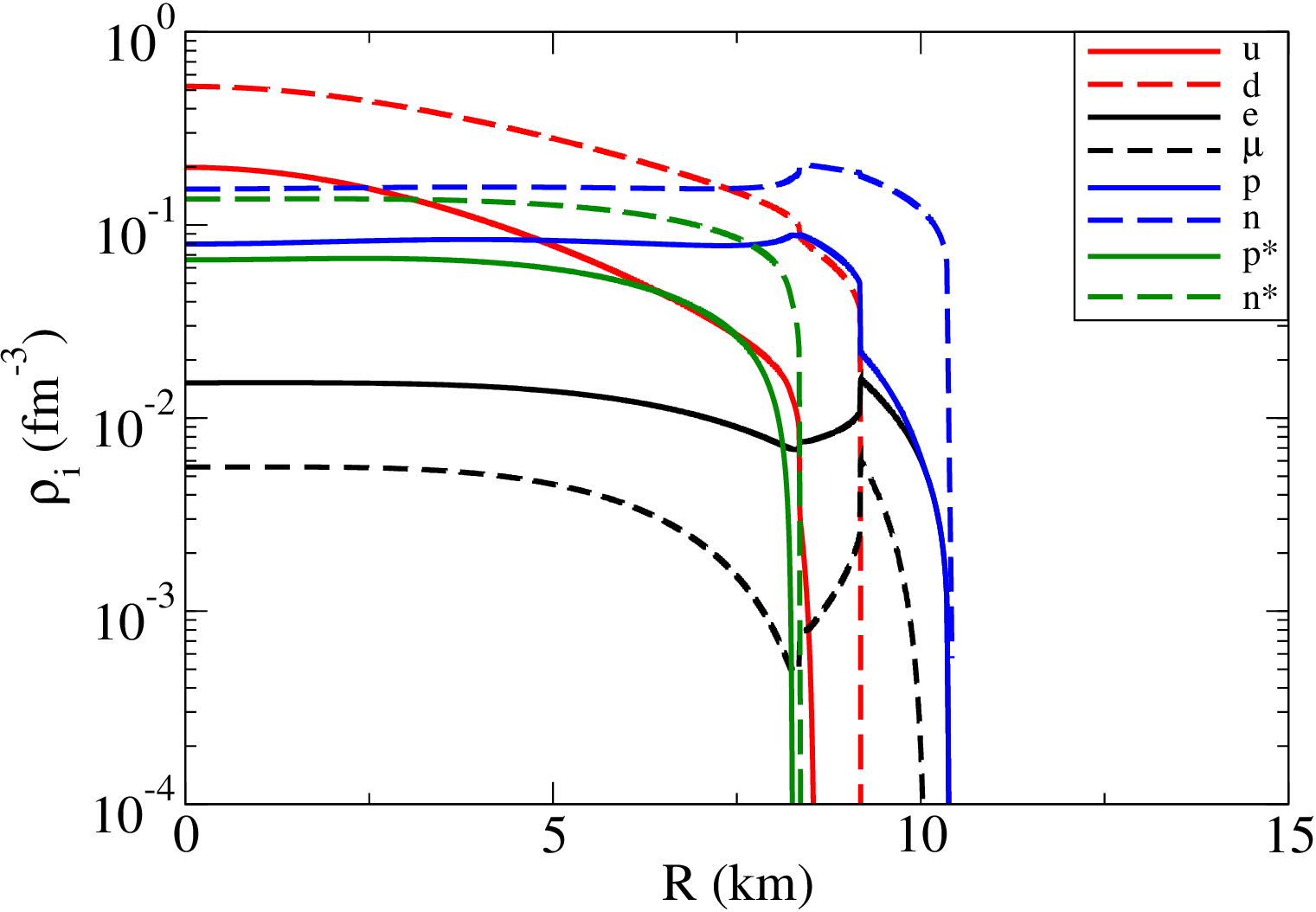}
 \caption{(Color online) Population (particle density) as a function of star
radius for a cold star using Model A. Quark densities are divided by $3$.}
 \label{popcold}
\end{figure}

By assuming chemical equilibrium and charge neutrality, we obtain an equation of
state that describes neutron star matter. This, in turn, can be used to obtain a
solution of the Tolman-Oppenheimer-Volkoff equation
\cite{Tolman:1939jz,Oppenheimer:1939ne}, which allows us to study how the system
behaves under the influence of gravity. We show in Fig.~\ref{mass} sequences of
stars with gravitational mass $M$, and radius $R$ for models A and B. For model
A, the maximum mass obtained for a cold star is $1.96$ $M_\odot$,
which is in agreement with observations of J1614-2230 (1.97~$\pm$ 0.04M$_\odot$)
\cite{Demorest:2010bx}. For a star with temperature of $30$ MeV, corresponding
to a proto-neutron star, the maximum mass obtained increases to $2.15$
$M_\odot$. The radius (corresponding to the most stable massive star) increases
from $10.70$ to $15.17$ km, the central baryonic density decreases from $1.16$
to $0.90$ fm$^{-3}$ and the central fraction of quarks decreases from $\sim
60\%$ to $\sim 50\%$ when a warm star is considered.

As for model B, the maximum mass obtained for a cold star reduces to $1.64$
$M_\odot$. For a  star with temperature of $30$ MeV the maximum mass obtained is
$1.75$  $M_\odot$. The radius of the most stable massive star, in this case,
increases  from $9.30$ to $14.28$ km, the central baryonic density decreases
from $1.58$  to $1.18$ fm$^{-3}$  and the central fraction of quarks decreases
from $\sim 50\%$ to $\sim 40\%$ when a warm star is considered. We clearly see
that the price paid to obtain a realistic compressibility is the reduction of
the maximum mass of the neutron star. This effect is generally model independent, but is known to be more pronounced in the parity model. It is important to keep in mind that we have not included in the present calculation any rotational or strong magnetic field effects in order to keep the effect of the appearance of the parity partners more clear. Both rotation and strong magnetic fields are known to increase the maximum mass of neutron stars \cite{Dexheimer:2008ax,Dexheimer:2011pz}.

\begin{figure}[t]
 \centering
\includegraphics[width=0.4\textwidth,clip,trim=8 0 0 0 ]{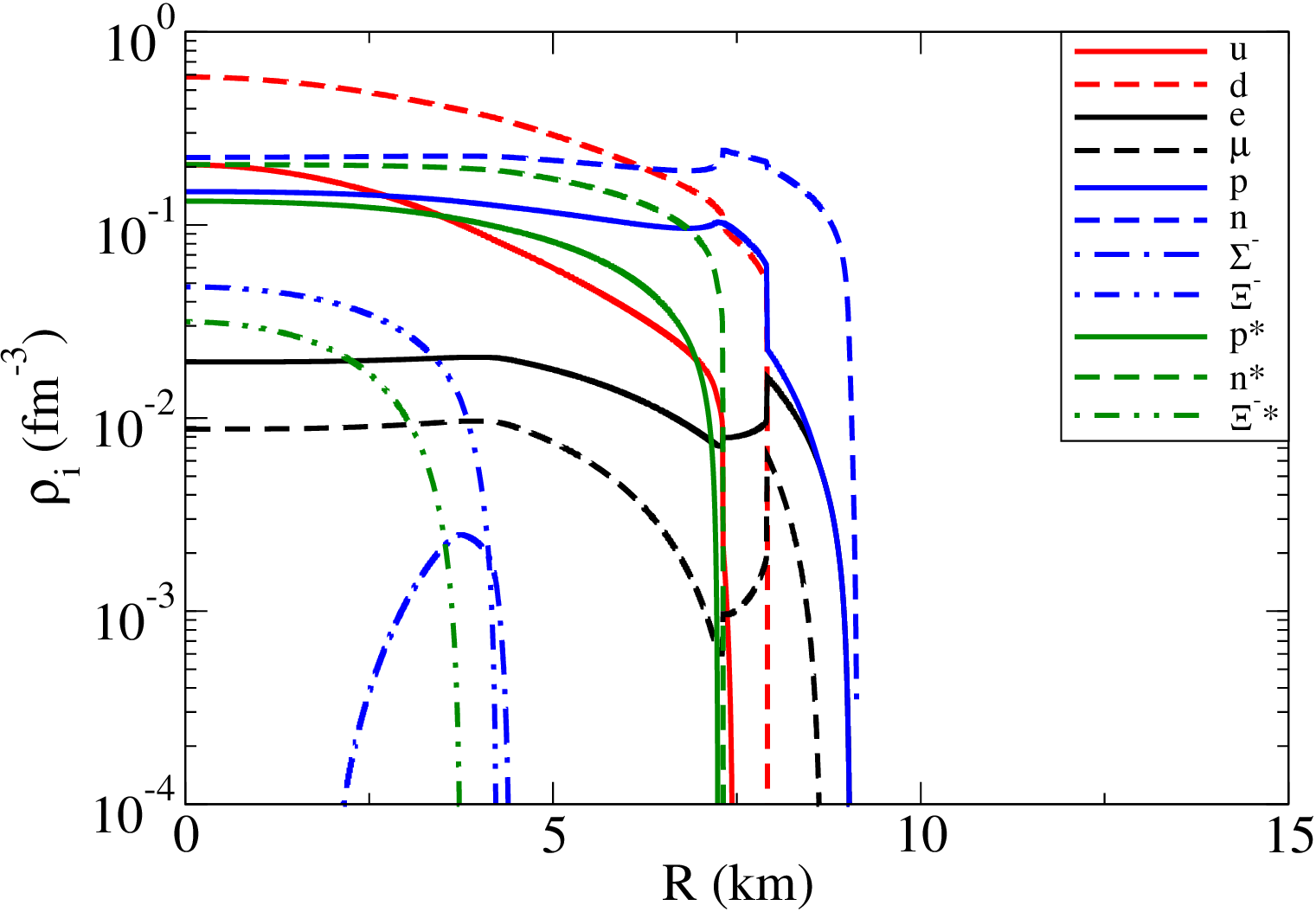}
 \caption{(Color online) Same as Fig.~\ref{popcold} but for model B.}
 \label{popcold2}
\end{figure}

In principle, the correct quantity to be used in order to analyze proto-neutron star temporal evolution is the entropy per baryon, instead of the temperature. Besides, trapped neutrinos should be included in earlier stages of star evolution, when the temperature (or entropy per baryon) is higher. Although these features are of great interest, our goal in this preliminary study of the topic is to analyze the application of the model to neutron stars in the simplest case of $T=0$ and the second simplest case of $T=30$ MeV. Nevertheless, we chose 30 MeV for the finite temperature case because it is approximately the maximum temperature reproduced in the center of neutron stars when the entropy per baryon is fixed to $2$. In this way, we study the two limiting cases of the star evolution ($T=0$ and $T=30$ MeV). Nevertheless, one can see that if the baryon number has to remain constant (like the case of isolated stars) not all stars from Fig.~\ref{mass} exist throughout the evolution, and some eventually collapse into black wholes. For example, the maximum baryon number for parametrization A is $2.73 \times 10^{57}$ at $T=0$ and $2.92 \times 10^{57}$ at $T=30$, and for parametrization B it is $2.24 \times 10^{57}$ at $T=0$ and $2.32 \times 10^{57}$ at $T=30$. In a detailed study of the matter, all the considerations above have to be taken into account.

\begin{figure}[t]
 \centering
\includegraphics[width=0.4\textwidth,clip,trim=8 0 0 0 ]{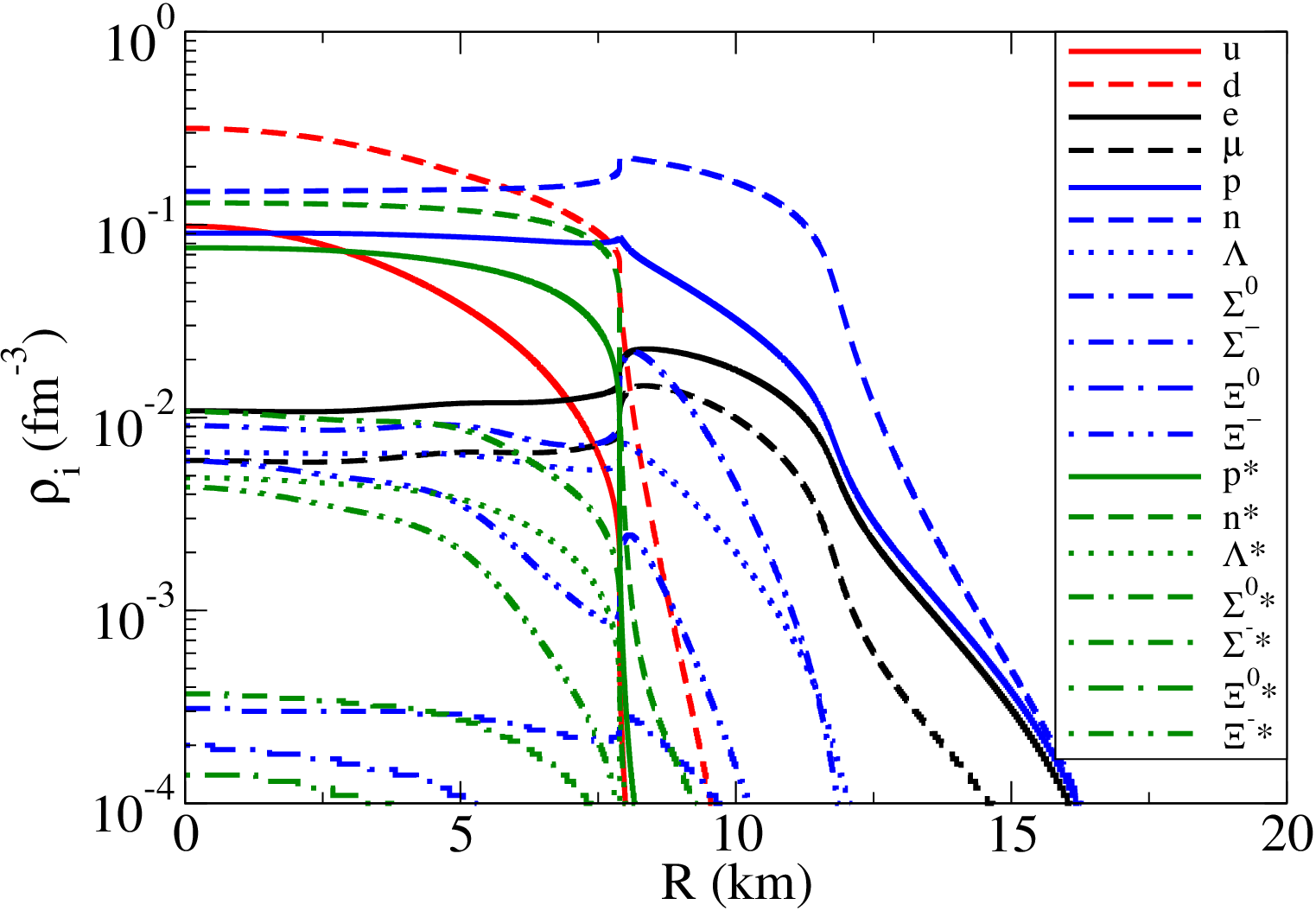}
 \caption{(Color online) Population (particle density) as a function of star
radius for a warm star $T=30$ MeV using Model A. Quark densities are divided by $3$.}
 \label{popwarm}
\end{figure}

The particle population of the cold maximum mass neutron stars obtained by both
models is shown in Figs.~\ref{popcold} and \ref{popcold2}. We see that at
low densities both stars consist only of neutrons, protons, electrons and muons.
Towards the center of the stars the quarks appear (first the $d$, followed by
the $u$) followed by the chiral partners (of neutrons and protons). The threshold for the quarks to appear is $1.56$ $\rho_0$ and $2.16$ $\rho_0$ for parametrizations A and B, respectively. It is
interesting to notice that in this model the baryons appear in small quantities
but do not
vanish at high densities. This is a consequence of the suppression of the
hadrons due to the excluded volume prescription. The hadrons are suppressed
exponentially by the quark pressure though not removed entirely. Note, however,
that their fraction
becomes so small that their contribution to the thermodynamics of the system is
negligible. Furthermore, at high densities, the amount of nucleons and the
respective chiral partners is very similar. Particles that contain strangeness,
despite being included in the model, do not appear or appear in very small
quantities in the cold stars. This is due their large bare masses and the slow
restoration of the strange scalar ($\zeta$) field to its zero value at high
densities.

\begin{figure}[t]
 \centering
\includegraphics[width=0.4\textwidth,clip,trim=8 0 0 0 ]{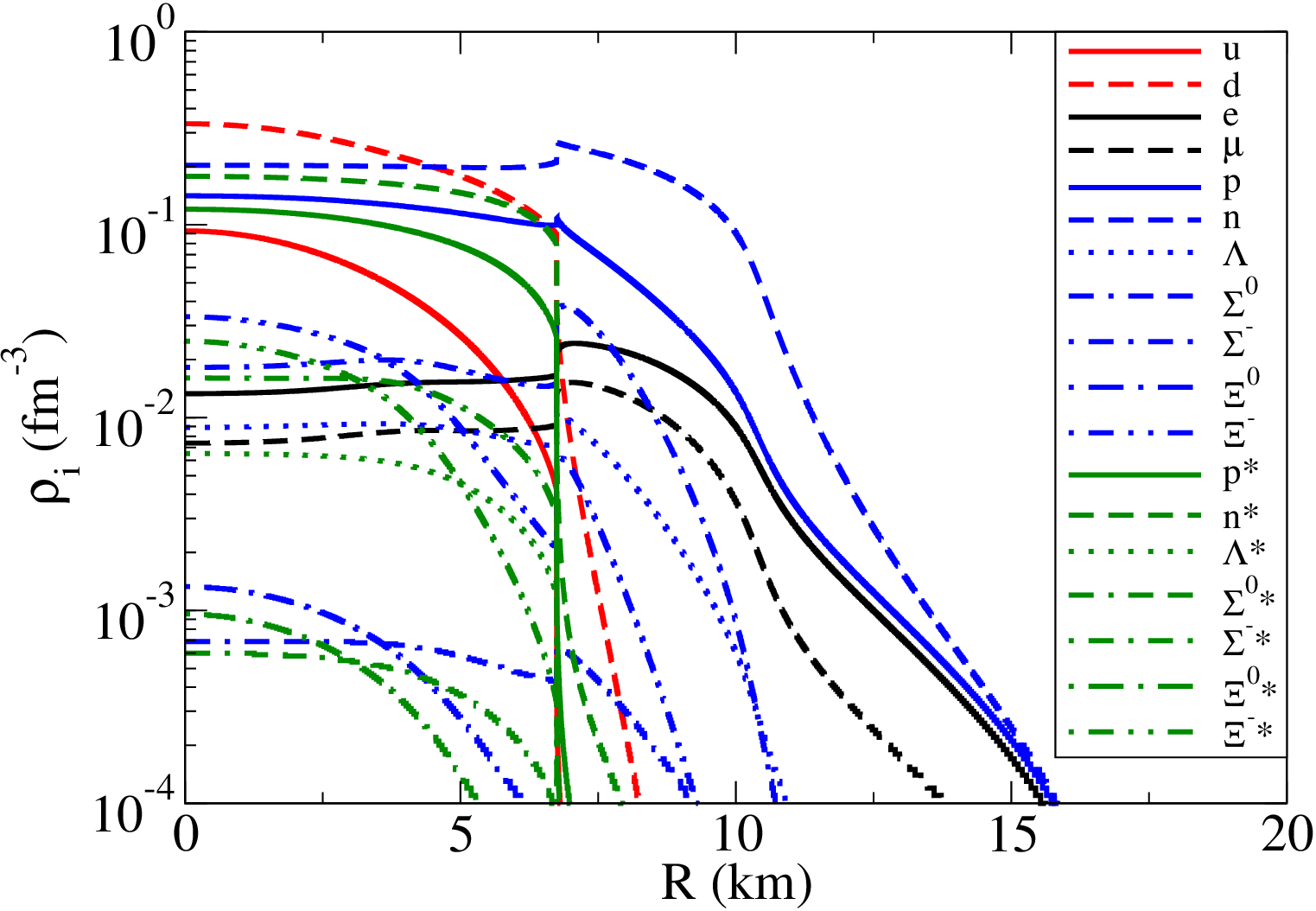}
 \caption{(Color online) Same as Fig.~\ref{popwarm} but for model B.} 
 \label{popwarm2}
\end{figure}

The decrease in central density stated above is not enough to prevent strange
particles to appear in the core of warm stars (Figs.~\ref{popwarm} and
\ref{popwarm2}). At low densities we see  $\Lambda$s, $\Sigma^-$s,
$\Xi^-$s and $\Sigma^0$s and at high densities $\Xi^0$s, all with their respective chiral partners. The strange quarks only appear at densities beyond the ones present in the studied stars. Overall, the amount of
particles that contain
strangeness is not high enough to significantly change star properties such as mass and
radius. These properties are influenced mainly by the pressure and energy
density changing substantially due to the high temperature, and the Polyakov
potential U, which only affects the system at finite temperature.\footnote{An example for
a Polyakov potential for the deconfinement order parameter that also depends on
baryon density, i.e. is applicable at $T=0$, can be found in
\cite{Dexheimer:2009hi}.}

Note that the order of the appearance of the hyperons in the star does not happen simply according to their vacuum masses. Their appearance is related to the relation between their effective masses and their effective chemical potentials. At zero temperature, for example, a particle only appears when its effective chemical potential is larger than its effective mass. For finite temperature, thermal effects also contribute making the above requirement more relaxed. But more importantly, the order of the particles in a star is strongly constrained by charged neutrality. This makes the appearance of negative charged hyperons more favorable, followed by charge neutral ones. Therefore, the only particles that appear in significant quantities for parametrization B at $T=30$ MeV are the $\Sigma^-$ followed by the $\Xi^-$ and respective chiral partners and they are the ones that still appear at $T=0$. For parametrization A, because there are no significant quantity of hyperons at $T=30$ MeV, there are also no hyperons in the $T=0$ MeV case.

For the cold case, there are basically two first order phase transitions in the
star, when the down quarks appear and when the up quarks and chiral partners
appear.  For the warm case, both of these transitions are
smoother and closer to each other. For the symmetric case, relevant for heavy
ion collisions, the phase transitions would be stronger for any temperature,
since chemical equilibrium and charge neutrality push the chiral symmetry
restoration (related to the appearance of chiral partners) to lower densities
with respect to the symmetric matter case and renders the transition smoother
\cite{Dexheimer:2008cv}.

\section{Thermal Evolution}

The thermal evolution of the stars was obtained by solving the full set of
equations that govern energy balance and transport in a relativistic star
\cite{Weber:1999qn}. We have considered the standard neutrino emission processes
for both quarks and hadrons, these are the direct Urca (DU), modified Urca (MU),
and Bremsstrahlung (BM) processes \cite{Yakovlev:2000jp,Iwamoto:1982zz}. The specific heat of the matter is given by the traditional
specific heat of relativistic fermions, as described in Ref. \cite{Page:2004fy}. The
thermal conductivity is calculated as described in
\cite{Flowers1981,Haensel:1991pi}. Finally, the boundary
condition that defines the surface temperature of the star is discussed in
Ref.~\cite{Gudmundsson1982,Gudmundsson1983,Page:2005fq,Blaschke:1999qx}. In our
calculations we assume a non-magnetized surface, with an accreted envelope of
$\Delta M/M = 1.0\times10^{-9}$.

\begin{figure}[t]
 \centering
\includegraphics[width=0.4\textwidth,clip,trim=3 0 0 0 ]{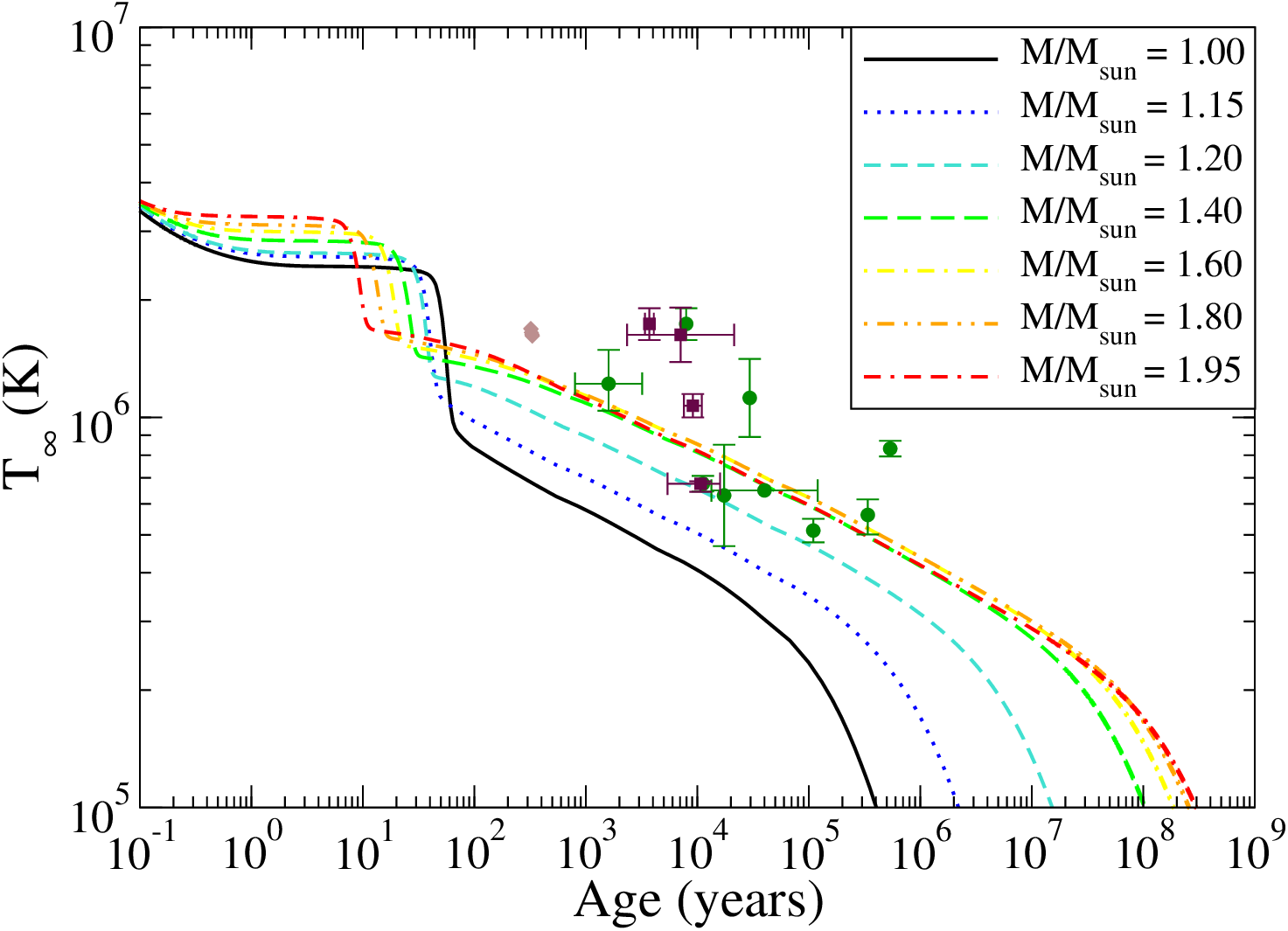}
 \caption{(Color online) Thermal evolution of stars of different masses for
model A.
$T_\infty$ denotes the redshifted temperature observed at infinity. The
observed data consists of circles denoting spin-down ages, squares kinematic
ages and diamonds the Cas A evolution.}
 \label{cool_delta_0}
\end{figure}

\begin{figure}[t]
 \centering
\includegraphics[width=0.4\textwidth,clip,trim=3 0 0 0
]{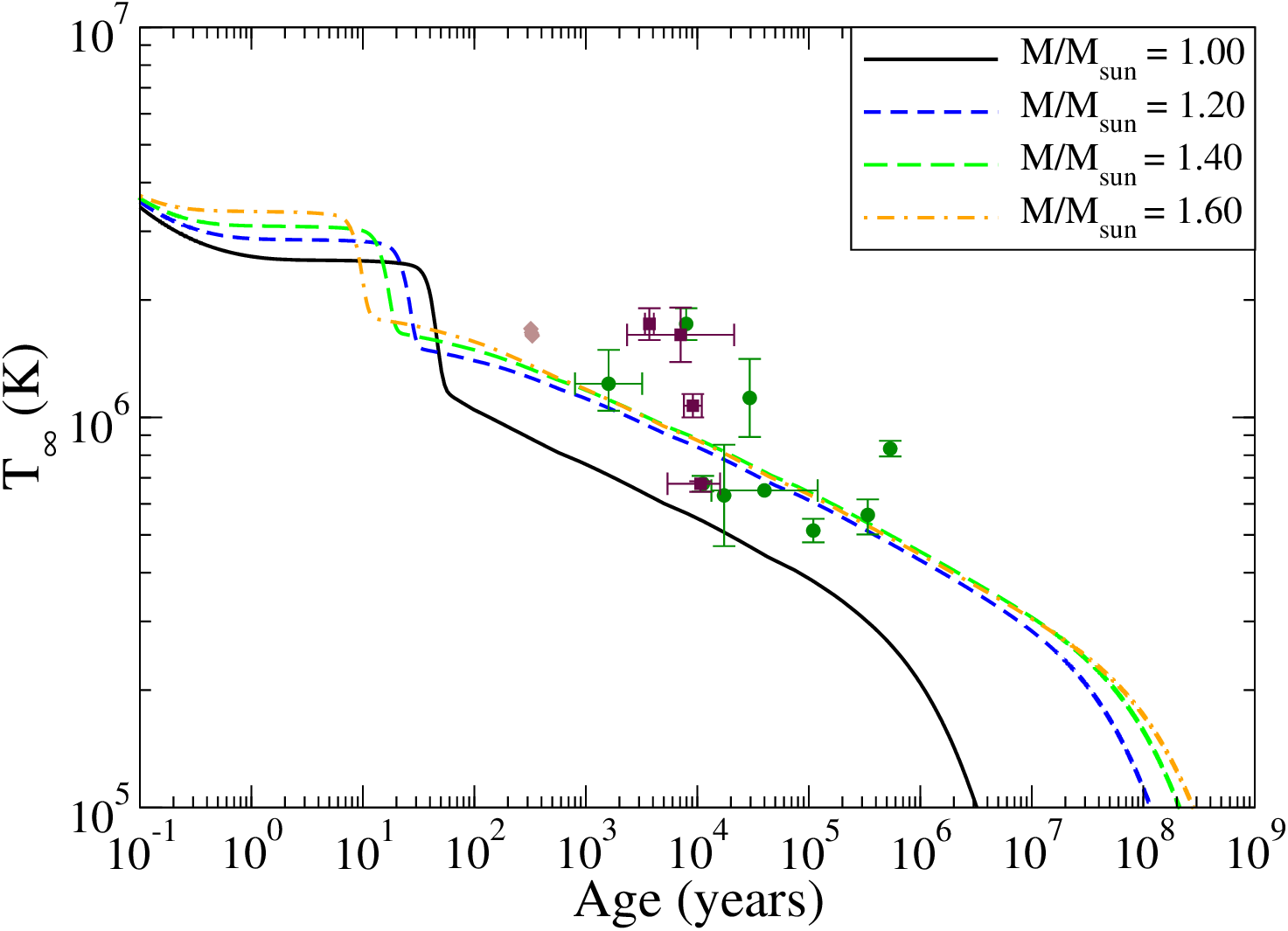}
 \caption{(Color online) Same as Fig.~\ref{cool_delta_0} but for model B.}
 \label{cool_delta_0_2}
\end{figure}

We have also taken into account neutron superfluidity when calculating the
thermal evolution. The pairing patterns considered were the neutron singlet
($^1S_0$) and triplet ($^3P_2$) states, as described in
\cite{Levenfish94,Schaab94}. We point out that it is possible that protons form
a singlet superconductor \cite{Page2011a,Yakovlev2011a},
however, the pairing of protons in the core is still not very well understood
(see \cite{Page2011b} and references therein). For that reason we consider only
neutron superfluidity at this stage.
We show in Figs.~\ref{cool_delta_0} and \ref{cool_delta_0_2} the cooling of
stars with different masses (for Models A and B, respectively), $T_\infty$ being the
redshifted surface temperature. We also plot a set of prominent observed cooling
data, where circles denote objects whose age was determined through spin-down,
squares through the object motion with respect to originating supernova
(kinematic age) \cite{Page:2004fy,Page:2009fu} and diamonds show the evolution
of Cas A \cite{Yakovlev2011a}.
We note that although our model exhibits temperatures and ages comparable
with those of Cas A,  it does not agree with the fast cooling exhibited by this
object in the last 10 years. We point out, however, that differently than
in references \cite{Page2011a} we have not removed fast neutrino emission
processes  by assuming that protons are in a superconducting
state in the entire core. As pointed out in \cite{Page2011b}, such stringent
condition is more easily achieved for low densities stars (less massive), since
theoretical models show that proton superconductivity does not extend to very
high densities. Under these conditions (no fast neutrino processes, and for the
critical temperature found in \cite{Page2011a}) our model might as well
reproduce the observed behavior of Cas A.

The results shown in
Fig.~\ref{cool_delta_0} clearly illustrate the effects of quark matter and the
chiral partners on the cooling of the star, which we discuss now. The low
density of the M/M$_\odot$ = 1.0 star forbids the presence of chiral partners.
Consequently the cooling of this object is qualitatively the same as that of
standard hybrid stars \cite{arXiv:1011.2233}. On the other hand, for stars with
higher central
densities, and thus higher
masses, we have chiral partner states being populated, and the cooling of the
star is substantially modified. Note that these stars are warmer than their
low-mass counterparts. Such behavior is the opposite of that of ordinary neutron
stars, which tend to exhibit a faster cooling for heavier objects. In the case
of ordinary stars the acceleration of the cooling with the increase of density
is connected to the increasing proton fraction. A higher proton fraction
throughout the star means a larger region where the DU process is present, and
thus a faster cooling. In our model, however, as the density increases
(equivalent to smaller radius in Fig.~\ref{popcold}) the proton fraction
decreases due to the increase of the chiral partners population. Because of
this, in our model, higher mass stars present a slower cooling. We also note
that the cooling of stars in model A and B are qualitatively equal, with the
difference that Model B cannot reproduce stars with masses higher than 1.64
M$_\odot$ (Fig.~\ref{cool_delta_0_2}).
Such an effect of slower cooling with increased mass has also been
observed in hybrid stars with a 
color-superconducting quark core \cite{Noda et al.(2011)}.

We should note here that these results should be considered with care. The
reason is that so far we have not considered neutrino processes
involving the chiral partners, which could certainly affect the cooling.  The
cross section of the processes involving the chiral partners have not been
determined so far, and are therefore excluded from this preliminary study.
If one shows that those cross sections are much smaller than the ones of
the traditional particles, i.e. the neutrino luminosity of this process is
substantially smaller than that of the traditional DU process; then the results
shown here would hold. These results can as well be easily tested, as soon as
there is enough statistical data on the relation between pulsar age, surface
temperature and mass.

The results of Fig.~\ref{cool_delta_0} and \ref{cool_delta_0_2} show that
the cooling of hybrid stars in our model is considerably slower than that of
traditional stars, even the one of the M/M$_\odot$ =
1.0 star that does not contain chiral partners. Additionally, as we have already
mentioned, we observe that
during the neutrino dominated era (ages $\lesssim 10^4$ years in ordinary
objects)  more massive stars exhibit a higher temperature. This substantially
delays the photon cooling era, that can be
identified by the bending of the cooling tracks at later times. We see that our
model, with the limitations discussed above, agrees fairly well with the cooling
data of the colder, and older objects, but fails to describe the temperature of
the very young and hot neutron stars.

\begin{figure}[t]
 \centering
\includegraphics[width=0.4\textwidth,clip,trim=3 0 0 0 ]{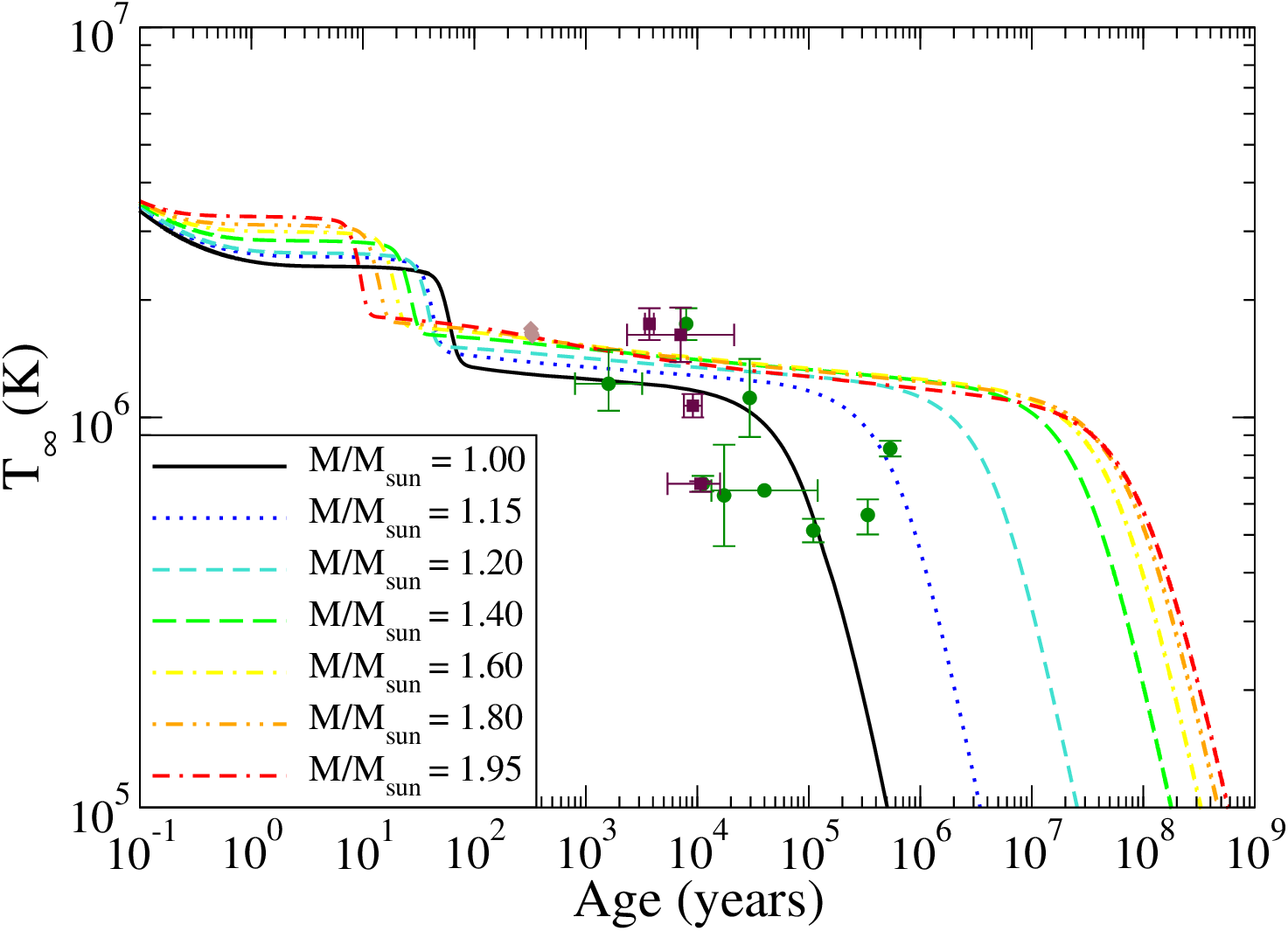}
 \caption{(Color online) Same as in Fig.~\ref{cool_delta_0} but for quark
matter paired with $\Delta = 10$ MeV.}
 \label{cool_delta_10}
\end{figure}

 Another possibility to be considered is quark pairing. Our study
considers a Color-Flavor-Locked (CFL) \cite{Alford:2007xm} {\emph like} phase,
where all quarks of all colors are paired. Note that since we consider small gap
values only ($\Delta =  10$ MeV), which are not expected to substantially affect
the EoS \cite{Alford:2003}, we do not include corrections in the EOS. In
this way, our analysis is still valid for any quark pairing scheme (not
necessarily color superconductivity), as long as it affects all quark flavors in
similar way. The quarks become a superconductor once the temperature falls below
the critical temperature $T_c$. For this study we assumed, like in reference
\cite{Blaschke:1999qx}, the value of $T_c = 0.4\Delta$. Once the
quarks become a superconductor the quark DU process is suppressed by a factor
$e^{-\Delta/T}$, and the modified Urca and the Bremsstrahlung process by a
factor $e^{-2\Delta/T}$. The specific heat of quark matter is also modified by a
factor
$3.2 (T_c/T)*(2.5 - 1.7(T/T_c) +3.6(T/T_c)^2)e^{-\Delta/T}$
\cite{Blaschke:1999qx}. 

We stress that the color superconducting phase
considered here is not, rigorously speaking, the traditional CFL phase. In the
CFL phase, all quarks of all flavors are paired. On another hand, in the 2
flavor superconducting state (2SC) one has the BCS pairing
between u and d quarks of colors r and g, which leaves some cooling channels
unsuppressed, thus leading to a faster cooling. Differently than the 2SC
phase, the model we consider in our calculations assumes that u and d quarks of
all colors are paired. As discussed above we consider only small gap values
($\Delta = 10$ MeV), which have little effect to the EoS. In that sense our pairing 
model is similar (but not exactly equal) to that discussed in
\cite{Blaschke2SCx}, labeled 2SC+x, where we have a 2SC pairing of the u and d
of two colors, and a residual pairing of the remaining unpaired quark, leading
to the effective pairing of all u and d quarks.

\begin{figure}
 \centering
\includegraphics[width=0.4\textwidth,clip,trim=3 0 0 0
]{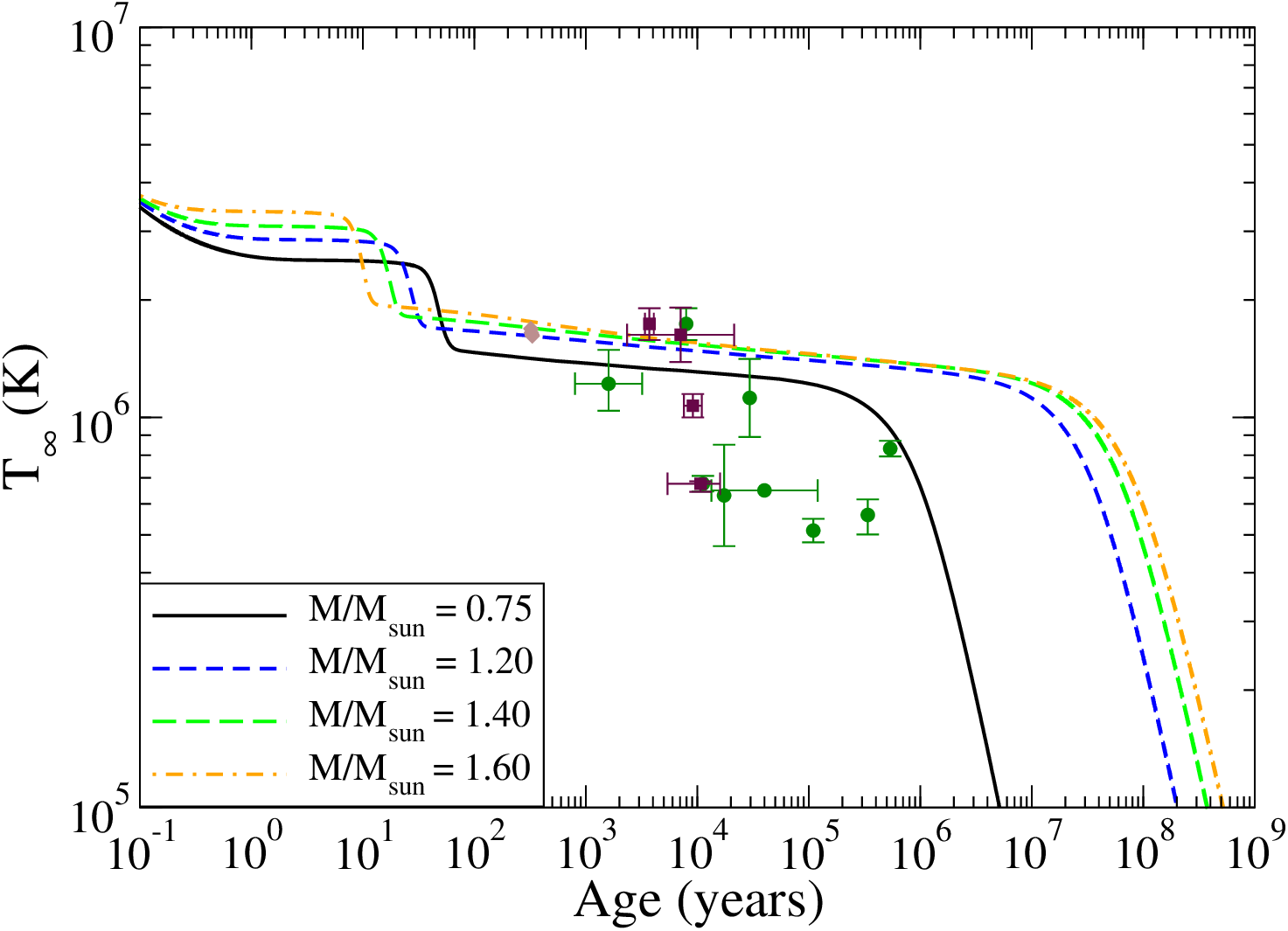}
 \caption{(Color online) Same as in Fig.~\ref{cool_delta_0_2} but for quark
matter paired with $\Delta = 10$ MeV.}
 \label{cool_delta_10_2}
\end{figure}

We show in Fig.~\ref{cool_delta_10} and \ref{cool_delta_10_2}  the cooling of
stars whose quark cores are in a superconductor state with $\Delta = 10$ MeV.
One can see that the suppression of the quark emission processes allow the stars
to be warmer during its neutrino cooling era being in better agreement with the
data of the very young and hot neutron stars. The time scale for the onset of
the photon cooling era is unchanged.  As was the case for stars with
unpaired quark matter, the cooling of stars in models A and B are very similar
As was the case before, we stress that these results should be
considered with great care, since the pairing scheme assumed here is
phenomenological. A more detailed study, taking into account microscopic effects
is warranted and will be performed in a future study.

\section{Conclusion}

We applied for the first time the $\rm SU(3)_f$ version of the parity doublet
model to cold, dense, charged neutral and chemically equilibrated matter. Note
that this approach is able to successfully describe the low density as well as
the high density regime of the QCD phase diagram, as it includes a
self-consistent deconfinement transition to quark matter. With these
ingredients, we can study for the first time the interplay between the baryon
octet, their respective chiral partners and quarks in neutron stars. The
chirally symmetric
but still confined matter obtained in Ref. \cite{Steinheimer:2011ea} disappears
when
charge neutrality and chemical equilibrium are taken into account. This is due
to
the early appearance of the down quark, that compensates the positive charge of
the proton. Such a partial appearance of the quarks was already suggested in
\cite{Blaschke:2008gd}. The up quark appearance happens, for both cold and warm
stars, approximately in the same region as the chiral partners (signaling the
chiral symmetry restoration). 

The zero temperature calculation using the same parametrization for the model
as the one calibrated for low and zero chemical potential (Model A) yields stars
with masses in agreement with the most massive observed pulsar (1.97~$\pm$
0.04M$_\odot$ \cite{Demorest:2010bx}). This parametrization, however, leads
to nuclear matter with high compressibility. On the other hand, parametrizations
leading to more realistic compressibility values (like model B) yield less
massive stars. As already shown in Ref. \cite{Dexheimer:2008cv}, corrections
that account for the baryonic Dirac sea effect such as the Relativistic Hartree
Approximation (RHA) can improve this situation. We note that the reason we did not consider any extra features that could possible improve the situation in this work is because we wanted to study in detail the population distribution in the star taking into account the relation between baryon chiral partners and quarks at high densities, since this had never been performed before. Work along this line is in
progress. The inclusion of finite temperature in the calculation allows more
massive stars in both of the cases studied, but still does not qualitatively
change the situation presented above.

We have performed cooling simulations for neutron stars whose microscopic
composition is given by our (cold) model. We have found that the presence of
the chiral partners affects the thermal evolution, effectively
suppressing the hadronic direct Urca process. This, in contrast to other models,
yields warmer stars during
the neutrino cooling era, and delays the onset of the photon cooling era.
Although we cannot effectively test such a prediction at the moment, we will be
able to do so, hopefully, in the near future.

We also considered the possibility of pairing in the quark phase, where a
CFL-like
pairing was assumed. The suppression of the quark neutrino emission processes,
brought on by pairing, reduces the total emission of neutrinos further, leading
to even warmer stars in the neutrino cooling era. We stress that the cooling
results presented here should be considered as a first approximation only,
since there are many factors that still need to be considered, like the neutrino
emissions from the chiral partners, and the effects of stronger quark pairing on
the EoS, which warrants future work along those lines. We believe, however, that
the cooling investigation put forth in this paper might be a good, qualitative
study of the cooling of neutron stars within a $\rm SU(3)_f$ doublet parity
model
composition, given that the available cooling data can be reproduced.
Furthermore, a recent work \cite{Negreiros:2012b} has shown that rotation might
have an important effect on the cooling of compact stars, thus we intend to
perform 2D simulations of the thermal evolution of stars described by our model.


\section*{Acknowledgments}

This work has been supported by GSI and the Hessian initiative for excellence
(LOEWE) through the Helmholtz International Center for FAIR (HIC for FAIR).
The computational resources were provided by the LOEWE Frankfurt Center for
Scientific Computing (LOEWE-CSC).
J.~S. acknowledges support by the Feodor Lynen program of the Alexander von
Humboldt foundation and from the Office of Nuclear Physics in the US Department
of Energy's Office of Science.
V.~D. acknowledges support from the Brazilian National Counsel of Technological
and Scientific Development (CNPq).

\end{document}